\documentclass[9pt, sigconf]{acmart}
\makeatletter
\usepackage{algorithmic}
\usepackage{graphicx}
\usepackage{textcomp}
\usepackage{makecell}
\usepackage{booktabs} % For formal tables
\usepackage{algorithm}
\usepackage{array} 
\usepackage{subcaption}
\usepackage[table]{xcolor}

\usepackage{rotating} % For text rotation
\usepackage{multirow}
\usepackage{tabularx}
\usepackage{diagbox}
\usepackage{caption}
\usepackage{footnote}
\usepackage{enumitem}
\usepackage{hyperref}
\usepackage{comment}
\def\BibTeX{{\rm B\kern-.05em{\sc i\kern-.025em b}\kern-.08em
    T\kern-.1667em\lower.7ex\hbox{E}\kern-.125emX}}
\begin{document}

% \title{Hardware-Aware Embedding Shaping: Joint \underline{C}ompression and \underline{Q}uantization for Robust Retrieval on \underline{CiM} (CQ-CiM)}
\title{CQ-CiM: Hardware-Aware Embedding Shaping for Robust CiM-Based Retrieval}

% CQ-CiM: Hardware-Aware Embedding Shaping for Robust CiM-Based Retrieval
% CQ-CiM: Hardware-Aware Compression and Quantization for Robust Low-Bit Embeddings on CiM
% CQ-CiM: Hardware-Aware Embedding Shaping for Robust Retrieval on CiM
% CQ-CiM: Low-Bit Embedding Shaping for Robust Retrieval on Compute-in-Memory
% CQ-CiM: Hardware-Aware Embedding Shaping for Efficient and Robust Retrieval on CiM
% CQ-CiM: A Unified Framework for Low-Bit Embedding Compression and Quantization on CiM
% \title{CQ-CiM: Towards Robust Retrieval on CiM via Joint \underline{C}ompression and \underline{Q}uantization Embedding}
% CQ-CiM: A Unified Framework for Joint Compression and Quantization in CiM-Based Retrieval

% \author{Authors \\ Authors \\ Institutes}

\author{Xinzhao Li*$^{1}$, Alptekin Vardar*$^{2}$, Franz Müller$^{2}$, Navya Goli$^{3}$, Umamaheswara Rao Tida$^{3}$, Kai Ni$^{4}$, \\ Xiaobo Sharon Hu$^{4}$, Thomas Kämpfe$^{2,5}$, Ruiyang Qin$^{1}$ \\ $^{1}$Villanova University, $^{2}$Fraunhofer IPMS, $^{3}$NDSU, $^{4}$University of Notre Dame, $^{5}$TU Braunschweig}
% \thanks{*Equal contribution.}
\renewcommand{\shortauthors}{Li and Vardar et al.}

% \thanks{* Equally contributed}

% \authornote{Equal contribution.}
% \vspace{-5ex}

% \author{Authors$^{1}$
% \\ $^{1}$Organization}

%% Rights management information.  This information is sent to you
%% when you complete the rights form.  These commands have SAMPLE
%% values in them; it is your responsibility as an author to replace
%% the commands and values with those provided to you when you
%% complete the rights form.
% \setcopyright{acmlicensed}
% \copyrightyear{2018}
% \acmYear{2018}
% \acmDOI{XXXXXXX.XXXXXXX}

%% These commands are for a PROCEEDINGS abstract or paper.
% \acmConference[Conference '26]{Make sure to enter the correct conference title from your rights confirmation emai}{June 03--05, 2018}{Woodstock, NY}
\acmConference[DAC `26]{Make sure to enter the correct conference title from your rights confirmation emai}{July 26--29, 2026}{Longbeach, CA}

%
%  Uncomment \acmBooktitle if th title of the proceedings is different
%  from ``Proceedings of ...''!
%
%\acmBooktitle{Woodstock '18: ACM Symposium on Neural Gaze Detection,
%  June 03--05, 2018, Woodstock, NY} 
% \acmISBN{978-1-4503-XXXX-X/18/06}

\begin{abstract}
Deploying Retrieval-Augmented Generation (RAG) on edge devices is in high demand, but is hindered by the latency of massive data movement and computation on traditional architectures. Compute-in-Memory (CiM) architectures address this bottleneck by performing vector search directly within their crossbar structure. However, CiM's adoption for RAG is limited by a fundamental ``representation gap,'' as high-precision, high-dimension embeddings are incompatible with CiM's low-precision, low-dimension array constraints. This gap is compounded by the diversity of CiM implementations (e.g., SRAM, ReRAM, FeFET), each with unique designs (e.g., 2-bit cells, 512x512 arrays). Consequently, RAG data must be naively reshaped to fit each target implementation.
Current data shaping methods handle dimension and precision disjointly, which degrades data fidelity. This not only negates the advantages of CiM for RAG but also confuses hardware designers, making it unclear if a failure is due to the circuit design or the degraded input data. As a result, CiM adoption remains limited. 
In this paper, we introduce CQ-CiM, a unified, hardware-aware data shaping framework that jointly learns \textbf{\underline{C}}ompression and \textbf{\underline{Q}}uantization to produce \textbf{\underline{CiM}}-compatible low-bit embeddings for diverse CiM designs. To the best of our knowledge, this is the first work to shape data for comprehensive CiM usage on RAG.

\end{abstract}

\maketitle

{\let\thefootnote\relax\footnotetext{* Equally contributed.}}
\vspace{-1ex}

\section{Introduction}
\label{sec:intro}
Retrieval-Augmented Generation (RAG) gives Large Language Models (LLMs) access to up-to-date knowledge. RAG is vital for on-device tasks, such as personalized AI or assistive robotics, where new real-time context is often outside the LLM's pretrained data. The core engine of RAG is a massive vector similarity search, often using Max Inner Product Search (MIPS) \cite{lewis2020retrieval} to find the most relevant information for a query. However, two main issues prevent wide RAG adoption on edge devices. First, limited edge resources cause large data movement between storage (e.g., SSD) and memory (RAM), leading to retrieval latency that can even exceed the LLM's inference time. Second, as data volume grows, the MIPS computation time on traditional systems scales poorly, creating a major performance bottleneck.

\begin{figure}[t]
    \centering
    \includegraphics[width=.9\linewidth]{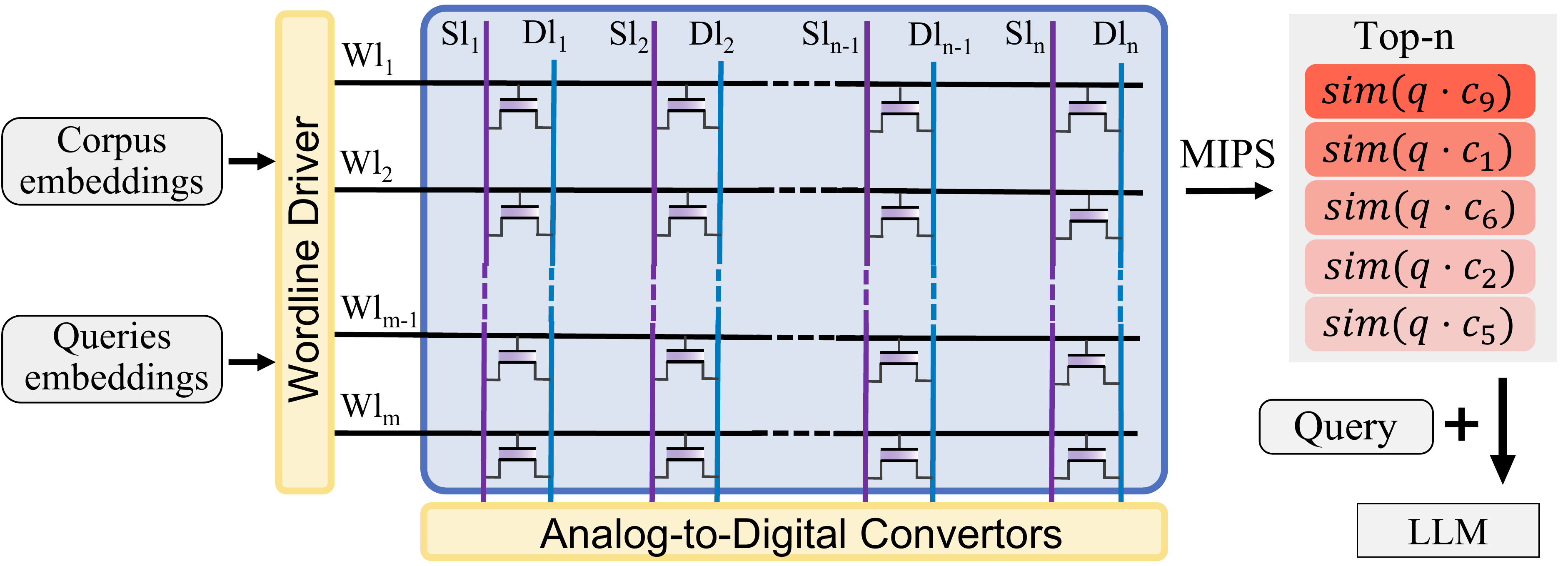}
    \vspace{-2.ex}
    % \caption{Illustration of RAG retrieval on a crossbar-based CiM architecture.}
    % \caption{Illustration of CiM-based embedding retrieval on a FeFET crossbar}
    % \caption{Illustration of CiM-based retrieval within a RAG pipeline on a crossbar array.}
    \caption{Illustration of CiM-based embedding retrieval on a FeFET crossbar array within a RAG pipline}
    \label{fig:rag_cim_architecture}
    \vspace{-6ex}
\end{figure}

% Retrieval-Augmented Generation (RAG) has become a cornerstone of modern AI, augmenting the capabilities of Large Language Models (LLMs) with external, up-to-date knowledge. The engine of this paradigm is efficient vector similarity search, a task that is increasingly critical for on-device applications in robotics and personalized AI, where smaller models are more reliant on retrieved knowledge. To meet the stringent latency and power demands of the edge, Compute-in-Memory (CiM) has emerged as a promising hardware architecture, leveraging its crossbar structure to perform vector search with exceptional energy efficiency.

Compute-in-Memory (CiM) emerges as a promising solution by directly addressing both RAG bottlenecks. The core MIPS operation (i.e., dot-product) is naturally suited to the CiM crossbar structure. As shown in Figure~\ref{fig:rag_cim_architecture}, the entire corpus of embedding vectors is stored as a matrix of conductances within the FeFET-based crossbar array. An input query vector is then applied to the array, performing a massive parallel vector-matrix multiplication directly in memory. This architecture fundamentally eliminates the data movement latency. It also solves the computational scaling problem, as the search latency is largely independent of the number of stored vectors. Therefore, CiM-based MIPS enables efficient on-device RAG by drastically reducing both latency and core resource usage.

Unfortunately, a fundamental ``representation gap'' compromises CiM's potential to accelerate RAG retrieval. On the one hand, sentence embedding models produce high-precision (e.g., FP32) and high-dimension (e.g., 1x2048) vectors by default. On the other hand, the physical CiM arrays can only support limited precision and dimension. Specifically, ReRAM-based CiM arrays sustain only 3 to 4 programable levels in controlled conditions but can degrade to an effective 2-level precision over time, while FeFET-based CiM arrays realistically support around 2 stable levels.
% Specifically, ReRAM-based CiM arrays can only store low-bit values (\textcolor{red}{[xx]}) in a single resistor, and FeFET-based CiM arrays can only store \textcolor{red}{[xx]}.
% are the physical realities of CiM hardware. CiM arrays impose two main constraints: a precision gap, supporting only low-bit values (e.g., 2-bit or 4-bit), and a dimension gap, having limited physical array sizes (e.g., 256x256). 
% Furthermore, these specific constraints are not fixed; they are highly diverse
The specific precision and dimensionality that can be supported vary across different CiM implementations (i.e., SRAM, ReRAM, and FeFETs). This hardware diversity means that embeddings must be "reshaped" on a per-target basis. The critical challenge is therefore the lack of a general, end-to-end method that can flexibly perform this shaping for any given CiM array.
% Unfortunately, a fundamental ``representation gap" compromises the CiM's potential to advance the information retrieval in RAG. During the information retrieval, the data sample is converted into a sentence embedding via the sentence embedding model. The precision and dimension, by default, can be FP32 and 1X2048. From the perspective of CiM, there are different implementations based on SRAM, ReRAM, or FeFET. For each different semiconductor devices, different settings can also be designed, such as 2-bit device, 3-bit device, based on the different device's aspect. Either setting can be much lower precision than the precision in default sentence embeddings. For the memory array, the size can also be vary, but normal would be smaller than the size of sentence embeddings as well. Hence, due to the various implementations and designs of CiM, the sentence embeddings are always required to be reshaped to fit into various CiM arrays. The thing is, there lacks of an end-to-end, generalizable method to shape the original sentence embeddings into CiM-friendly settings. 

To bridge this gap, we introduce an unified hardware-aware, parameter-efficient framework to joint compression and quantization for CiM's information retrieval and relevant downstream tasks like RAG, referred to as CQ-CiM, that shatters this bottleneck by simultaneously shaping both precision and dimension to fit any given crossbar-based CiM design. Our approach introduces an unified, end-to-end trainable pipeline built on three synergistic components: (1) a parameter-efficient adapter for lightweight fine-tuning of the given embedding model; (2) a learnable compression head to reduce dimensionality to fit CiM array size constraints; and (3) a non-linear quantization head to map embeddings to low-bit CiM-compatible format. This entire framework is trained using a specifically designed loss function that uniquely enables self-supervised learning by leveraging a contrastive objective, removing the restriction of data labeling. This contrastive loss is complemented by a reconstruction loss (MSE) to guide and stabilize the compression and quantization heads.
% This is co-designed with a reconstruction loss (MSE) to guide and stabilize the compression and quantization heads. 
% This integrated framework, which unifies \textbf{\underline{C}}ompression and \textbf{\underline{Q}}uantization for \textbf{\underline{CiM}}, is hereby introduced as \textbf{\underline{CQ-CiM}}. 
Furthermore, our design uniquely allows physical hardware characteristics, such as device variance (noise), to be incorporated into the training process (via noise injection), shaping the embeddings to be robust to such noise.

% In this work, we propose a hardware-aware, parameter-efficient framework that shatters this bottleneck by learning to simultaneously and flexibly shape both the precision and dimension of embeddings for CiM-based retrieval. Our approach introduces a lightweight, learnable adapter, which is fine-tuned using a multi-objective loss function. This function uniquely incorporates not only task performance and desired dimensionality but also the physical characteristics of the target hardware, such as device variance. The adapter features two synergistic heads: a Quantization Head that leverages non-linear functions to map embeddings to low-bit representations (e.g., 1.58-bit) suitable for analog storage on 2-bit devices, and a Compression Head that reduces the embedding dimension.

Our major contributions are summarized as follows:
\begin{itemize}
    \item To the best of our knowledge, this is the first work to introduce a unified framework that jointly handles precision and dimension shaping with explicit hardware-awareness, enabling embedding representations to be directly aligned with CiM crossbar array designs.
    \item We introduce a novel adaptation pipeline that incorporates dimension compression, precision quantization, and parameter-efficient tuning into a single process. This design enables a single embedding model to be flexibly specialized for a wide range of CiM devices with different hardware constraints.
    \item We demonstrate our framework's superiority through comprehensive experiments, showing it significantly outperforms SOTA methods on diverse retrieval benchmarks, and maintains high robustness in end-to-end RAG evaluations under practical CiM device variations 
    % realistic hardware noise.
\end{itemize}

\section{Background and Challenge}
\newcolumntype{Y}{>{\centering\arraybackslash}X}

\begin{table}[t!]
\fontsize{7pt}{5.pt}\selectfont

% \footnotesize
    \centering
    % \caption{Device non-ideality modeling for different real and synthesized devices. For devices with more than two levels, the device variation for each level is depicted as $L_x$. }
    % \caption{Device variation profiles (D-1 to D-5) used to model state-dependent conductance noise in multi-level memory cells. $\sigma_v$ denotes the per-level standard deviation of each nominal state.}
    % \caption{State-dependent device variation profiles used for constructing the conductance transition model. Each multi-level device is represented by nominal conductance states $L_0$–$L_3$ (corresponding to the four levels in a 2-bit cell), and $\sigma_v$ denotes the standard deviation of the Gaussian variation applied to each level.}
    \caption{Level-dependent device variation profiles used in the conductance transition model. For multi-level cells, $L_0$–$L_3$ denote the nominal states and $\sigma_v$ their corresponding Gaussian deviations.}

    \vspace{-3ex}
    \begin{tabularx}{\columnwidth}{c*{5}{Y}}        
    \toprule
        \multirow{2}{*}{Name} & \multirow{2}{*}{\# of Levels}  & \multicolumn{4}{c}{Device Variations $\sigma_v$} \\
                  &   & $L_0$ & $L_1$ & $L_2$ & $L_3$ \\
        \midrule
        $RRAM_1$ (Device-1)  & 1 & 0.0100 & 0.0100 & 0.0100 & 0.0100\\
        $FeFET_2$ (Device-2) & 4 & 0.0067 & 0.0135 & 0.0135 & 0.0067\\
        $FeFET_3$ (Device-3) & 4 & 0.0049 & 0.0146 & 0.0146 & 0.0049\\
        $RRAM_4$ (Device-4)  & 4 & 0.0038 & 0.0151 & 0.0151 & 0.0038\\
        $FeFET_6$ (Device-5) & 4 & 0.0026 & 0.0155 & 0.0155 & 0.0026\\
        \bottomrule
    \end{tabularx}
    \vspace{-9ex}
    \label{tab:var}
\end{table}
% \vspace{0.1cm}

\subsection{Analog CiM Arrays and Multi-Level FeFETs}
\label{sec:crossbar_nvm}

CiM architectures have been realized using several device technologies. SRAM-based CiM is the most mature and widely adopted due to its high reliability and negligible device variation \cite{sram}. Meanwhile, non-volatile memory (NVM) technologies such as resistive RAM (ReRAM) \cite{rram} and ferroelectric transistor (FeFET) devices \cite{soliman2023first} can also be used to build crossbar-based CiM arrays that support analog current-mode vector–matrix multiplication (VMM) and multi-level weight storage. Our CQ-CiM framework is designed to support all three CiM device classes.

\begin{figure}[!h]
    \centering
    \includegraphics[width=\linewidth]{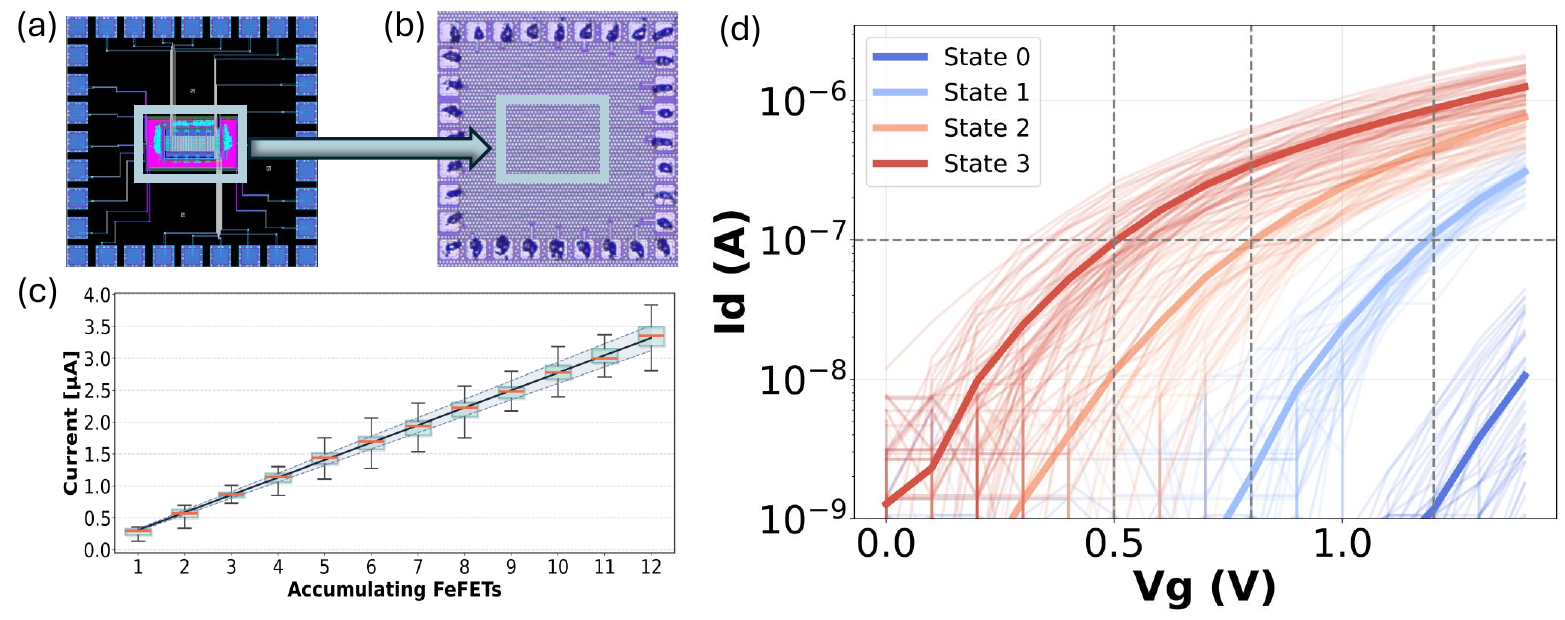}
    \vspace{-6ex}
    \caption{Experimental characterization and functionality of the FeFET-based compute-in-memory array in 28\,nm CMOS. 
    (a) Layout of the mixed-signal FeFET crossbar array
    (b) fabricated die photograph, 
    (c) measured bitline current accumulation under progressive wordline activation, and 
    (d) programmed 2-bit multi-level FeFET operation of 50 devices exhibiting four distinct $V_\mathrm{T}$ states ($L_0$–$L_3$).}
    \label{fig:fefet_CiM}
   \vspace{-4.ex}
\end{figure}

Crossbar-based CiM arrays perform analog VMM by programming device conductances at WL-BL intersections and accumulating bitline currents according to Ohm's and Kirchhoff's laws. This structure naturally fits similarity search operations such as MIPS and enable large parallelism in minimal data movement. Figure~\ref{fig:fefet_CiM}(a-c) shows the 28nm GF HKMG mixed-signal FeFET CiM array \cite{tsa, qubo} used as the reference platform, where measured bitline currents exhibit near-linear accumulation where activating multiple wordlines \cite{fefet_acc}, confirming the array's suitability for analog inference.

To support larger embedding capacity under tight area and latency constraints, per-cell storage density becomes essential. Multi-level FeFETs provide an efficient solution by storing multiple quantized values within a single device through stable intermediate threshold-voltage states \cite{soliman2023first} . As shown in Figure~\ref{fig:fefet_CiM}(d), the fabricated 2-bit FeFET cells demonstrate four well-separated $L_0$ to $L_3$ states\cite{f1,f2,f3,f4}.  Measurements from 50 fabricated devices validate these states, showing well-separated $I_\mathrm{D}$–$V_\mathrm{G}$ characteristics with minimal overlap, as illustrated in Fig.~\ref{fig:fefet_CiM}(d).

Integrating multi-level cells into CiM arrays provides a natural match to low-bit (2-bit) neural representations but introduces level-dependent conductance variation \cite{jjap}. Table~\ref{tab:var} summarizes the variation profiles ($L_0$–$L_3$) used in this work, each modeled by a nominal conductance and a Gaussian deviation $\sigma_v$. These parameters are derived from three representative measured devices—two RRAM technologies~\cite{yao2020fully, 
liu2023architecture} and one FeFET device~\cite{wei2022switching}—denoted as $RRAM_1$, $RRAM_4$, and $FeFET_2$. We further extrapolate two synthesized variants ($FeFET_3$, $FeFET_6$) to span a broader multi-level range. An 
$x$-level device supports $x$ distinct conductance states (e.g., $\sigma_{L_2}=0.01$ indicates a variation of 0.01 at level 2), forming the basis of the transition model used in our hardware-aware training.

\subsection{Representation Gap between RAG and CiM}

In RAG, a pretrained sentence embedding model \cite{qin2025empirical} converts both queries and corpus documents into high-dimensional floating-point embeddings, and retrieval is performed through Max Inner Product Search (MIPS). Compute-in-Memory (CiM) architectures provide an appealing substrate for this operation: embedding vectors can be stored as crossbar conductances, and queries can be applied as analog voltages to perform massively parallel vector–matrix multiplication (VMM), yielding low-latency and low-movement similarity search \cite{qin2024robust}. 

However, deploying RAG on CiM reveals a fundamental \emph{representation gap}. Modern sentence embeddings are high-precision (FP32) and high-dimensional (e.g., 768–2048) \cite{reimers2019sentence}, whereas CiM arrays—whether based on SRAM, ReRAM, or FeFET—support only a few multi-level conductance states (typically 1–2 bits per cell) and fixed array dimensions (e.g., 64$\times$64 or 128$\times$128) \cite{soliman2023first, lee2022multi}. Bridging this mismatch requires reshaping embeddings along two orthogonal axes: compressing their dimension to fit array size and quantizing their values to match low-bit device states. Naive PCA-based compression \cite{jolliffe2016principal} or uniform quantization \cite{jacob2018quantization} degrades semantic structure and ignores device non-idealities such as multi-level variance and read noise \cite{shim2020two}, making them insufficient for real CiM deployment.

% \textbf{Limitations of prior CiM-based embedding works.} 
Existing CiM accelerators such as ISAAC, PRIME, and Newton demonstrate efficient analog VMM for similarity search \cite{shafiee2016isaac, chi2016prime, he2020newton}, but all assume that embeddings are already stored in a CiM-compatible form and do not address how high-dimensional FP embeddings should be reshaped before storage. Retrieval quantization methods \cite{jegou2010product} and recent low-bit embedding schemes \cite{jeong20244bit} improve semantic efficiency but remain hardware-agnostic, overlooking array dimension constraints \cite{qin2025nvcim}, multi-level programming behavior \cite{ni2019novel}, and device variation \cite{ni2020impact}. 
% Even CiM-focused optimizations primarily target circuit-level VMM efficiency or non-ideality compensation \cite{jain2019cxdnn}, rather than co-designing the embedding representation itself. 
Even CiM-focused optimizations primarily target circuit-level VMM efficiency or non-ideality compensation \cite{jain2019cxdnn}, without addressing how the embedding representation itself should be reshaped for CiM constraints.
Consequently, no prior work jointly addresses dimensionality compression, low-bit quantization, and device-aware robustness in a unified, end-to-end framework tailored for CiM—leaving a critical gap between RAG’s embedding requirements and the constraints of practical CiM hardware.

\section{Proposed Work}
% \textcolor{red}{[Could be removed if out of space]} In this section, we present the design of our framework, CQ-CiM. We begin by elaborating its overall architecture and data flow, which is shown in Figure~\ref{fig:framework}. Then we describe the key components of this pipeline in sequence: the parameter-efficient adaption and compression strategy, the hardware-aware noise injection method, and the non-uniform quantization head for low-bit mapping. Finally, we present the dual-objective loss function that unifies the end-to-end training process. 

% After the training process, as shown in the right part of Figure~\ref{fig:framework}, the trained framework--CQ-CiM, can be used to generate the CiM-compatible embeddings, enabling the usage of CiM in RAG.

\begin{figure}[t!]
    \centering
    \includegraphics[width=0.65\linewidth]{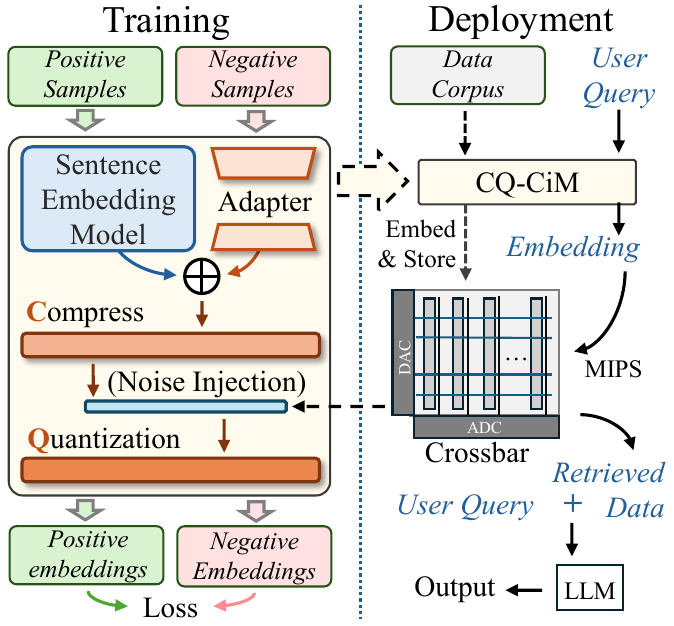}
    \vspace{-3ex}
    % \caption{Overview of the proposed hardware-aware representation learning framework. }
    \caption{Overview of \textbf{CQ-CiM}. \textbf{Left:} The framework jointly shapes embedding dimension and precision via a LoRA-based adapter with noise injection for hardware robustness. \textbf{Right:} Demonstrate that the trained embedding model based on our framework (CQ-CiM) is used to bridge the CiM and RAG.} 
    \vspace{-4ex}
    \label{fig:framework}
\end{figure}

\subsection{Overall Architecture}

% This section introduces the full CQ-CiM framework, including its overall architecture, PEFT-based adaptation, compression module, noise injection, and low-bit quantization design. The overall architecture of CQ-CiM is illustrated in Figure~\ref{fig:framework}. CQ-CiM enables an end-to-end training pipeline to convert high-precision and high-dimension sentence embeddings into a low-precision and low-dimension format compatible with various designs and implementations of CiM crossbar arrays.
This section introduces the full CQ-CiM framework, including its overall architecture, adaptation module, compression module, noise injection, and low-bit quantization design. The overall architecture of CQ-CiM is illustrated in Figure~\ref{fig:framework}. CQ-CiM enables an end-to-end training pipeline to convert high-precision and high-dimension sentence embeddings into a low-precision and low-dimension format compatible with various designs and implementations of CiM crossbar arrays.

The data flow passes through three main stages in the training pipeline. 
First, the input sentence is encoded by the backbone sentence embedding model equipped with a LoRA adapter \cite{hu2021loralowrankadaptationlarge}, used as our parameter-efficient adaptation mechanism. LoRA injects a lightweight low-rank update into selected weight matrices, allowing the encoder to adjust its representations without full fine-tuning. This enables the embedding model to adapt to our downstream compression and quantization objectives while keeping the trainable parameter budget small.
Then, the output is fed into the compression head to reduce its dimensionality based on the crossbar array size (e.g., 128D). Finally, the compressed embeddings pass through our noise injection module to be injected with simulated device noise. The processed embeddings are then quantized by our designed quantization head to map them to low-bit values.

To enable the training pipeline in a self-supervised manner, positive and negative samples are constructed by modifying the dropout rate of the embedding model \cite{qin2024robust}. As shown in Figure~\ref{fig:framework}, after these two types of samples pass through the pipeline, the loss is calculated and used for backpropagation. The Deployment side of the figure illustrates the deployment process: the CQ-CiM block (containing the LoRA, compression, and quantization heads) shapes the corpus and query embeddings before they are sent to the CiM crossbar for efficient RAG retrieval.

% \subsection{Parameter-Efficient Adaptation and Compression Head}
\subsection{Adaptive Embedding Compression}

The sentence embedding model is the core of the framework. The way to get the embedding model involved in the training is critical. Freezing the model can be computationally efficient, but it potentially limits the framework's learning space, as the backbone cannot adapt its representations to various compression and quantization tasks. Conversely, fine-tuning all parameters of the embedding model risks creating a parameter imbalance \cite{houlsby2019parameter}, where the massive model's gradients overwhelm the learning of lightweight compression and quantization heads. 
% Therefore, we employ the PEFT approach based on LoRA. 
Therefore, we employ a LoRA-based adaptation mechanism.
Backed by the experiments in the following section, this choice provides an optimal balance, allowing the core embedding model to adapt while keeping training costs low and the parameter space balanced.

A similar design rationale applies to the compression head. We evaluated several alternatives, including PCA \cite{Mackiewicz1993PCA} and autoencoder \cite{bank2021autoencoders}. PCA is a strong baseline, but it is a non-learnable method and therefore cannot be optimized end-to-end or adapt to the task-specific loss functions. An Autoencoder is learnable, but it is incompatible with our single-stage, unified training objective, as it typically requires a separate pre-training phase before its decoder is discarded. In this work, we employ a simple yet effective dense projection layer as our compression head. This design is fully learnable, computationally lightweight, and perfectly compatible with our end-to-end pipeline.

% Collaboratively training the sentence embedding model along with the compression. Compared with freezing the sentence embedding model, enabling the adaptation of the sentence embedding model, along with the training of compression and quantization heads, can largely enlarge the learning space of the entire framework. Meanwhile, if we enable the full-parameters of the sentence embedding model to train, we can cause an imbalance of parameters between the embedding model and CQ heads, not mention the large cost of training. Instead, we add parameter efficient adapter to the sentence embedding model, so we can leverage the sentence embedding model adaptation and the training of CQ heads. 

% Along with the training of the sentence embedding model, we have our compression head. There are several ways to design, such as autoencoder or PCA. For the autoencoder, the problem is that is is incompatiable with the end-to-end compression and quantziation training. The autoencoder must be trained first, removing the decoder, and train with quantization head. For PCA, it's not a learnable method, cannot adapt the new querys well. In this work, we employ a simple and effective dense head as our compression head.

\subsection{Noise Injection Training}

The purpose of our noise injection training is to make the final embeddings resilient to hardware-level device variance, when non-volatile memories are applied in CiM. The placement of this module is a critical design choice, as shown in Figure~\ref{fig:framework}. We inject the noise directly after the compression head, but before the quantization head. This location is chosen to balance the noise impact: if noise were injected before compression, its effect would be diminished by the projection layer; conversely, if noise were injected after quantization, it would become a discrete "bit-flip" problem. This would cause significant information distortion and, crucially, prevent the quantizer's learnable thresholds from adapting to the noise.
The device variation parameters in Table~\ref{tab:var} are derived from measured FeFET and ReRAM arrays, ensuring that the simulated noise closely reflects real CiM hardware behavior.

To implement this, we take the compressed embedding, $emb_C \in \mathbb{R}^d$, where the dimension $d$ matches the compression setting. We construct a noise mask, $\boldsymbol{\epsilon}$, which is a new tensor with the same dimension $d$ as the compressed embedding.

To build this mask, we use the user-provided hardware characteristics from Table~\ref{tab:var}. For $K$ conductance levels, the user defines $K-1$ static thresholds $\{\tau_1, ..., \tau_{K-1}\}$ (e.g., $\{0.25, 0.5, 0.75\}$) and $K$ variance values $\{L_0, ..., L_{K-1}\}$. 
Each element $\epsilon_j$ in the noise mask is associated with a level index $k = \text{find\_level}(emb_{C,j}; \{\tau_k\})$, and is drawn from a Gaussian distribution with level-dependent deviation:
% Each element $\epsilon_j$ in the noise mask is then drawn from a Gaussian distribution whose standard deviation $\sigma_k$ is determined by the level $k$ that the corresponding data $emb_{C,j}$ falls into:
% \begin{equation}
%     \text{Let } k = \text{find\_level}(emb_{C,j} \text{ using static } \{\tau_k\})
% \end{equation}
\begin{equation}
    \epsilon_j \sim \mathcal{N}(0, L_k^2) \quad \text{where } \sigma_k = L_k \text{(refer to Table~\ref{tab:var})}
\end{equation}
Finally, this level-aware noise mask $\boldsymbol{\epsilon}$ is scaled by a global noise factor $\sigma_g$ (simulating the overall disturbance level) and added to the compressed embedding to produce the final, noise-injected embedding $emb_{\sigma}$:
\begin{equation}
\label{eq:noise_injection}
    emb_{\sigma} = emb_C + \sigma_g \cdot \boldsymbol{\epsilon}
\end{equation}
This noise-injected embedding $emb_{\sigma}$ is then passed to the subsequent quantization head for training.

\subsection{Quantization Head and Joint Loss}

Low-bit quantization is essential for deploying embeddings on CiM crossbar arrays, which are often composed of memory cells that support only a few discrete conductance levels (i.e., 2-bit). A key challenge is that the distribution of these compressed embeddings is typically highly non-uniform. Instead of employing the nonuniform quantization method \cite{choukroun2019low}, we formalize compressed embeddings into a matrix and employ the Nonuniform-to-Uniform Quantization (N2UQ) method \cite{gholami2021surveyquantizationmethodsefficient, liu2022nonuniformtouniformquantizationaccuratequantization}, which is originally designed to quantize model weights, to lower the precision of the matrix.

\begin{figure}[t]
\centering
\includegraphics[width=1.\linewidth]{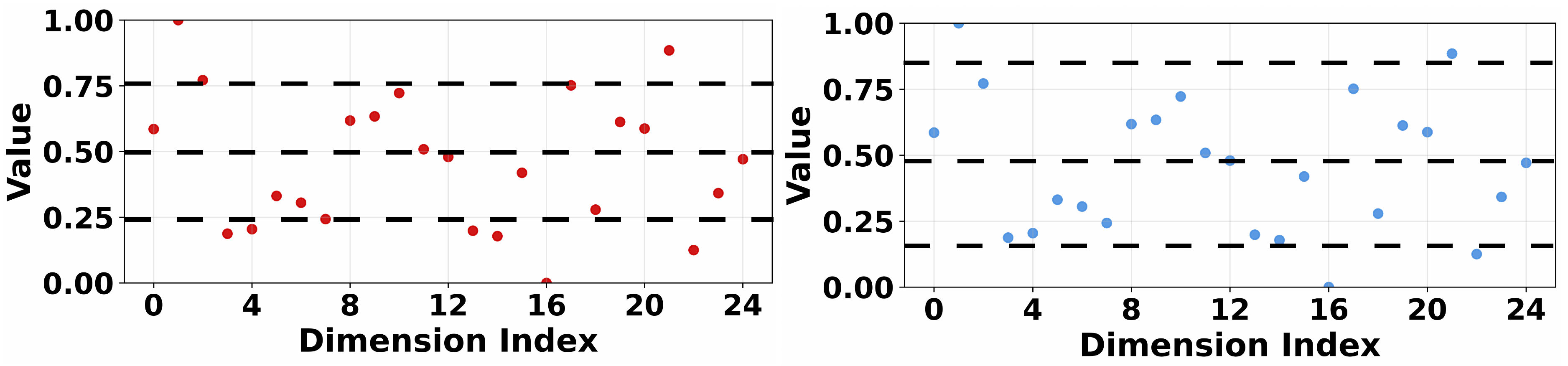}
\vspace{-6ex}
\caption{Visualization of quantized values of a single embedding (dimension=25) using fixed nonuniform threshold (left) and learned N2UQ threshold (right)}
\vspace{-5ex}
\label{fig: quant comprare}
\end{figure}

To ensure the N2UQ head learns effectively in conjunction with the compression and noise injection modules, we employ a joint loss function. Our framework is trained end-to-end using a dual-objective loss that combines contrastive learning with reconstruction-based supervision. This design reflects the practical constraint of corpus-only training: contrastive learning refines the semantic geometry, while reconstruction stabilizes the compression and quantization modules. As shown in Figure~\ref{fig: quant comprare}, fixed nonuniform quantization (left) fails to capture the true data distribution, while N2UQ (right) learns to distribute the embedding values more effectively across the available levels. 

Following the behavior of the quantization head, we next detail the training objectives that make the quantization and compression heads learnable within a unified pipeline.

\textbf{Contrastive Loss:}
For self-supervised learning, we leverage a contrastive objective. Two dropout-masked views \cite{gao2021simcse} of the same sentence are processed by our framework to form a positive pair $(h_i, h_i^+)$, while all other batch samples serve as negatives:
\begin{equation}
\mathcal{L}_{\mathrm{CSE}}
= -\log 
\frac{
e^{\mathrm{sim}(h_i,h_i^{+})/\tau}
}{
\sum_{j=1}^{N}
\left(
e^{\mathrm{sim}(h_i,h_j^{+})/\tau} 
+
e^{\mathrm{sim}(h_i,h_j^{-})/\tau}
\right)
}.
\end{equation}
We use this objective primarily to update the LoRA adapter to enhance the quality of the learned representations.
% to refine the semantic space.

\textbf{Reconstruction Loss:} To guide the compression and quantization heads, we use a standard mean squared error (MSE) reconstruction objective, which minimizes the distance between the original floating-point embedding and the final compressed-quantized output:
\begin{equation}
\mathcal{L}_{\text{MSE}}
= \bigl\| f_{\text{orig}}(x) - 
\hat{f}_{\text{cq}}(x) \bigr\|_2^2.
\end{equation}
Although the output embedding has lower dimensionality and discrete precision, this reconstruction objective stabilizes training and helps preserve the quality of the compressed and quantized embeddings.

\textbf{Overall Objective:} The final loss is a weighted sum of the two objectives:
\begin{equation}
\mathcal{L}
= \mathcal{L}_{\text{CSE}}
+ \lambda_{\text{MSE}} \mathcal{L}_{\text{MSE}},
\end{equation}
where $\lambda_{\text{MSE}}$ balances the semantic and reconstruction goals. In practice, for 384$\rightarrow$128 compression, a moderately large weight ($\lambda_{\text{MSE}} \approx 5$–$10$) is effective.

\section{Experimental Evaluation}
In this section, we present experiments to evaluate the performance, design, generalizability, and hardware-aware robustness of our proposed CQ-CiM framework. Our primary task is information retrieval, while RAG serves as the downstream validation of robustness. We first present our main results, benchmarking our complete, optimized framework against existing baselines across five diverse information retrieval datasets to demonstrate its advanced performance in the low-bit range. Then, we perform a detailed ablation study to justify our framework's advantage, validating its design. Later on, we validate the generalizability of our framework on four different sentence embedding models. Finally, we provide the end-to-end evaluation on RAG, the downstreaming task, where the device variances from non-volatile memories are injected to assess the robustness of CQ-CiM on different CiM implementations. By default, experiments are run on an NVIDIA A10 GPU.

% In this section, we present a comprehensive evaluation designed to justify the structure of our experimental methodology. Since CQ-CiM addresses both representation learning and hardware-level constraints, a standard “setup → results’’ format is insufficient. Instead, we evaluate the framework along four complementary dimensions: (1) low-bit retrieval performance to validate overall effectiveness, (2) ablations to isolate the contribution of each design choice, (3) generalization across multiple embedding backbones, and (4) end-to-end RAG robustness under realistic device variation. Together, these components provide a complete assessment of both the algorithmic and hardware-aware aspects of CQ-CiM.

% In this section, we design our experimental evaluation to validate the performance, design, generalizability, and hardware-aware robustness of our proposed CQ-CiM framework. we evaluate the framework along four complementary dimensions: low-bit retrieval performance to validate overall effectiveness,  ablations to isolate the contribution of each design choice, generalization across multiple embedding backbones, and end-to-end RAG robustness under realistic device variation. Together, these components provide a complete assessment of both the algorithmic and hardware-aware aspects of CQ-CiM.

\subsection{Evaluating CQ-CiM Performance}
\label{sec:4.1}
To validate our framework's performance, we benchmark it across 5 informative retrieval datasets: ARCChallenge \cite{wachsmuth2018retrieval}, NanoHotpotQA \cite{yang2018hotpotqa}, CQADupStackGisRetrieval \cite{hoogeveen2015}, and ArguAna \cite{wachsmuth2018retrieval}. We use \textit{all-MiniLM-L6-v2} \cite{reimers-2020-multilingual-sentence-bert} as the base sentence embedding model with the embedding dimension set to 128. Our framework is compared with three baselines: UMAP \cite{mcinnes2020umapuniformmanifoldapproximation}, PCA + uniform quantization \cite{Mackiewicz1993PCA}, and native scale (Vanilla) \cite{qwen3embedding}. 

\begin{table*}[t]
\centering
% \small
\fontsize{7pt}{4pt}\selectfont
\caption{
Comparison of different compression approaches at 128D (128-dimensional embeddings) across five retrieval datasets. Corpus sizes are shown in parentheses (e.g., 9.35k = 9.35k documents). \
% Best performance for each bit setting is in bold. 
PQ-8bit is included as a high-precision vector-quantization reference (not CiM-compatible) and does not necessarily represent the highest retrieval performance.
% PQ-8bit is included only as a high-precision vector-quantization reference (not CiM-compatible). 
% Because PQ uses codebook-based quantization rather than per-dimension similarity preservation or task-level optimization, its retrieval accuracy may be lower than that of our learned low-bit CiM-friendly embeddings.
}
\vspace{-3ex}
\label{tab:bit_comparison}
\resizebox{\linewidth}{!}{
\begin{tabular}{c|c|cc|cc|cc|cc|cc}
\toprule
\multirow{2}{*}{Precision} &
\multirow{2}{*}{Method} &
\multicolumn{2}{c|}{ARCChallenge (9.35k)} &
\multicolumn{2}{c|}{NanoHotpotQA (5.09k)} &
\multicolumn{2}{c|}{CQADupStack (37.6k)} &
\multicolumn{2}{c|}{FiQA (57.6k)} &
\multicolumn{2}{c}{ArguAna (8.67k)} \\
& & Recall@5 & nDCG@10 & Recall@5 & nDCG@10 & Recall@5 & nDCG@10 & Recall@5 & nDCG@10 & Recall@5 & nDCG@10 \\ 
\midrule

\multirow{3}{*}{1 bit}
& UMAP    & 0.024 & 0.021 & 0.160 & 0.151 & 0.035 & 0.033 & 0.026 & 0.026 & 0.167 & 0.138 \\
& PCA     & 0.069 & 0.059 & 0.490 & 0.450 & 0.145 & 0.246 & 0.192 & 0.194 & 0.410 & 0.329 \\
& Vanilla & 0.071 & 0.055 & 0.310  & 0.357 & 0.255 & 0.227 & 0.182 & 0.179 & 0.378 & 0.307 \\
& \textbf{CQ-CiM} & \textbf{0.082} & \textbf{0.062} & 0.380 & 0.393 & \textbf{0.302} & \textbf{0.256} & \textbf{0.212} & \textbf{0.210} & \textbf{0.449} & \textbf{0.360} \\
\midrule

\multirow{3}{*}{1.58 bit}
& UMAP    & 0.010 & 0.009 & 0.080 & 0.070 & 0.003 & 0.003 & 0.013 & 0.012 & 0.078 & 0.070 \\
& PCA     & 0.098 & 0.076 & 0.470 & 0.501 & 0.201 & 0.311 & 0.293 & 0.292 & 0.511 & 0.418 \\
& Vanilla & 0.094 & 0.073 & 0.530 & 0.463 & 0.345 & 0.297 & 0.240 & 0.248 & 0.474 & 0.386\\
& \textbf{CQ-CiM} & 0.095 & \textbf{0.077} & \textbf{0.540} & \textbf{0.543} & \textbf{0.382} & \textbf{0.322} & \textbf{0.288} & \textbf{0.303} & 0.507 & \textbf{0.419} \\
\midrule

\multirow{3}{*}{2 bit}
& UMAP    & 0.033 & 0.027 & 0.140 & 0.157 & 0.068 & 0.056 & 0.036 & 0.042 & 0.230 & 0.190 \\
& PCA     & 0.083 & 0.068 & 0.480 & 0.482 & 0.192 & 0.309 & 0.244 & 0.249 & 0.410 & 0.338 \\
& Vanilla & 0.092 & 0.071 & 0.470 & 0.470 & 0.346 & 0.302 & 0.244 & 0.248 & 0.468 & 0.385 \\
& \textbf{CQ-CiM} & \textbf{0.110} & \textbf{0.082} & \textbf{0.540} & \textbf{0.547} & \textbf{0.399} & \textbf{0.346} & \textbf{0.299} & \textbf{0.297} & \textbf{0.529} & \textbf{0.433} \\
\midrule

PQ (8-bit)
& \textbf{Reference} & 0.117 & 0.074 & 0.760 & 0.475 & 0.334 & 0.241 & 0.435 & 0.222 & 0.456 & 0.244 \\

\bottomrule
\end{tabular}
}
\vspace{-2ex}
\end{table*}

Under the dimension of 128, we evaluate three low-bit settings: 1-bit \cite{wang2023bitnet}, 1.58-bit \cite{ma2024era1bitllmslarge}, and 2-bit \cite{chee2024quip2bitquantizationlarge}. Such low-bit precision levels are commonly adopted in crossbar-based CiM designs \cite{soliman2023first, lee2022multi}. 
In addition, we include the uncompressed PQ-8bit \cite{jegou2010product} as a high-precision reference. Note that PQ is a codebook-based vector quantization method rather than a task-adapted or CiM-oriented representation, and therefore its performance does not necessarily exceed that of optimized low-bit embeddings. 
% In addition, we include the uncompressed PQ-8bit \cite{jegou2010product} as a high-precision reference. However, PQ-8bit is incompatible with CiM’s dot-product computation pattern, as PQ operates on codebooks rather than per-dimension conductance values. 
All performance is measured using Recall@5 and nDCG@10 \cite{wang2013theoretical}.
% Under the dimension of 128, we evaluate three low-bit settings: 1-bit \cite{wang2023bitnet}, 1.58-bit \cite{ma2024era1bitllmslarge}, and 2-bit \cite{chee2024quip2bitquantizationlarge}. They are widely used in crossbar array design \cite{soliman2023first, lee2022multi}. In addition, we also include the uncompressed PQ-8bit \cite{jegou2010product}, using as a high-precision reference. For PQ-8bit, it is incompatible with the CiM’s dot-product computation pattern, as PQ operates on codebooks rather than per-dimension conductance values. All performance is measured using Recall@5 and nDCG@10 \cite{wang2013theoretical}.
% Experiments run on a NVIDIA A10.

As shown in Table~\ref{tab:bit_comparison}, the results validate our approach. The UMAP baseline's performance collapses at ultra-low-bit settings, as its manifold structures are not robust to aggressive discretization. PCA performs competitively at 1.58-bit but becomes unstable at 1-bit. The Vanilla truncation method consistently underperforms, confirming that this simple approach discards important semantic information. In contrast, our CQ-CiM framework consistently achieves the best or most competitive performance across all datasets, showing its largest advantages at the most aggressive 1-bit and 2-bit settings. UMAP performs poorly because its non-linear manifold is highly sensitive to aggressive low-bit quantization, making it inherently unsuitable for CiM-style discretized representations.

% As a concluding remark, these results demonstrate the value of our joint tuning, compression, and quantization approach. By jointly optimizing the components, our framework produces robust, semantically consistent low-bit embeddings that outperform existing isolated compression methods under realistic CiM hardware constraints.

% As a concluding remark, these results highlight the value of jointly learning adaptation, compression, and quantization. Within this unified framework, our method produces robust, semantically consistent low-bit embeddings that outperform isolated compression approaches under realistic CiM hardware constraints.

As a concluding remark, these results highlight the value of jointly learning adaptation, compression, and quantization. Within this unified framework, our method produces robust, high-quality low-bit embeddings that outperform individual embedding-shaping approaches under realistic CiM hardware constraints.

\subsection{Validating Core Components}
We perform an ablation study to examine how each learnable component contributes to the overall effectiveness of CQ-CiM. Specifically, we examine three learnable components that shape embeddings: the adaptation mechanism, the compression head, and the quantization head. All experiments are conducted on the ArguAna dataset using the \textit{all-MiniLM-L6-v2} encoder, where the original 384D FP32 embeddings are shaped to 128D 2-bit for evaluation.

Across the three components, we compare the case where LoRA is involved with two settings ($rank=4, \alpha=8$ and $rank=8, \alpha=16$) and the case where LoRA is not used to enable the embedding model adaptation along with the training process. Then, we compare our Dense projection layer with autoencoder (AE). Finally, we compare the fixed 2-bit Straight-Through-Estimator (STE) and the learned N2UQ method. 

\begin{figure}[t!]
    \centering
    \includegraphics[width=0.8\linewidth]{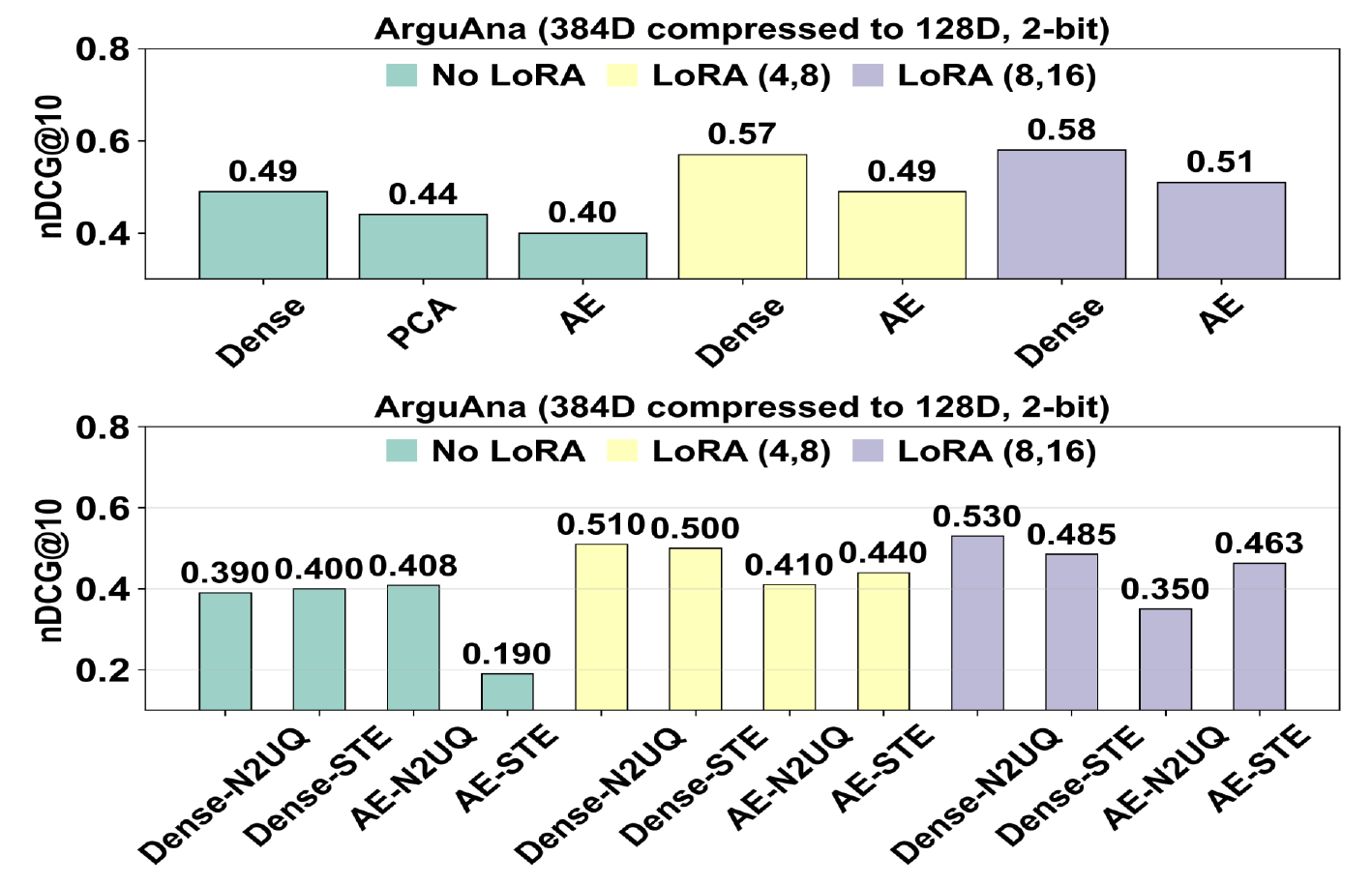}
    \vspace{-2ex}
    % \caption{Overview of the proposed hardware-aware representation learning framework. }
    \caption{Ablation study on ArguAna. Retrieval performance across LoRA settings, compression methods, and quantization strategies. LoRA (8,16) + Dense + N2UQ achieves the strongest retrieval accuracy.}
    \vspace{-5ex}
    \label{fig:ablation_bar}
\end{figure}
% Experiments run on a NVIDIA A10.

% \begin{figure}[t!]
%     \centering
%     \includegraphics[width=.8\linewidth]{figures/figure2_combined.pdf}
%     \vspace{-2ex}
%     % \caption{Overview of the proposed hardware-aware representation learning framework. }
%     % \caption{Retrieval accuracy curves for four embedding models across varying embedding dimensions under four quantization precisions (1-bit, 1.58-bit, 2-bit, and INT4), illustrating how performance changes as dimension and precision vary.}
%         % The curves illustrate how performance scales as both dimension and precision are reduced.
%     \caption{Retrieval accuracy curves for four embedding models across different embedding dimensions and four quantization precisions (1-bit, 1.58-bit, 2-bit, and INT4).}

%     \vspace{-8ex}
%     \label{fig:summary}
% \end{figure}

\begin{figure*}[t!]
    \centering
    \includegraphics[width=.85\textwidth]{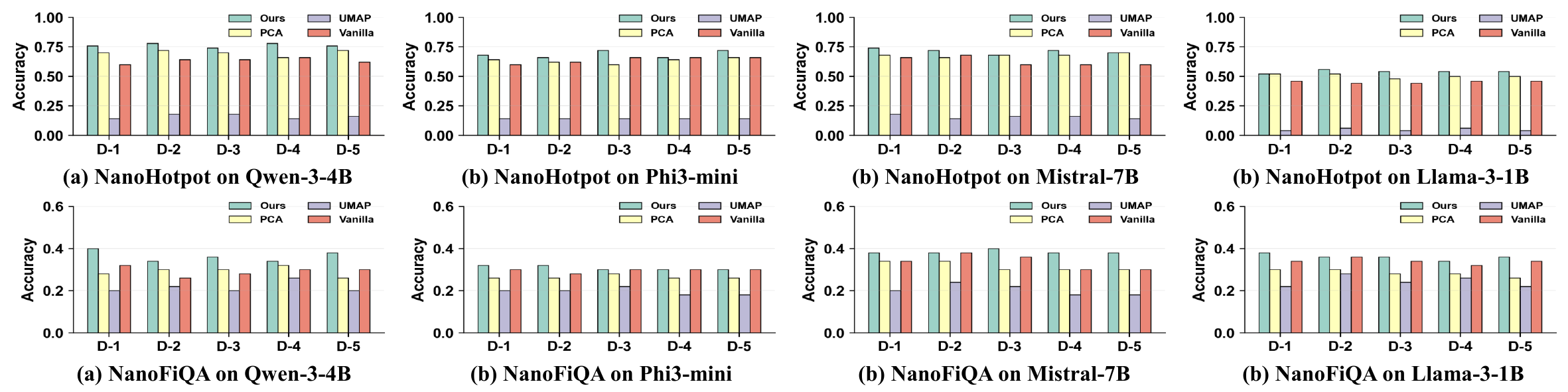}
    \vspace{-2ex}
    \caption{End-to-end RAG accuracy under device non-idealities. D-1 to D-5 correspond to the device variation settings in Table~\ref{tab:var}. Quantized embeddings are perturbed using transition matrices derived from these device variations.}
    \vspace{-2ex}
    \label{fig:rag-llm-device}
\end{figure*}

As shown in Figure~\ref{fig:ablation_bar}, the results validate our design choices. First, enabling adaptation with LoRA consistently improves performance over the No-LoRA baseline, with the $r=8,\alpha=16$ setting achieving the strongest gains. Second, the Dense projection proves more robust than the autoencoder variants, which degrade retrieval quality due to their mismatch with the single-stage training pipeline. Finally, N2UQ provides the largest individual improvement, with its learned thresholds outperforming the fixed STE by 3–7\% absolute across all settings. 

Taken together, these observations indicate that the combination of LoRA adaptation, Dense compression, and N2UQ quantization forms the most effective embedding-shaping pipeline, which is the setting used for the evaluations in Section~\ref{sec:4.1}.

\captionsetup[subfigure]{justification=centering,singlelinecheck=false}
% \subsection{Scaling Behavior of low bits embeddings across models}
\subsection{Generalizability Across Embedding Models}

% \begin{figure}[t!]
%     \centering
%     \begin{subfigure}[t]{0.48\linewidth}
%         \centering
%         \includegraphics[width=\linewidth]{figures/Nomic-embed-text-v1.5.png}
        
%         \label{fig:nomic}
%     \end{subfigure}
%     \hfill
%     \begin{subfigure}[t]{0.48\linewidth}
%         \centering
%         \includegraphics[width=\linewidth]{figures/KaLM-embedding-multilingual-mini-instruct-v2.5.png}
        
%         \label{fig:kalm}
%     \end{subfigure}

%     \vspace{4pt}

%     \begin{subfigure}[t]{0.48\linewidth}
%         \centering
%         \includegraphics[width=\linewidth]{figures/Qwen3-Embedding-0.6B.png}
        
%         \label{fig:qwen}
%     \end{subfigure}
%     \begin{subfigure}[t]{0.48\linewidth}
%         \centering
%         \includegraphics[width=\linewidth]{figures/Granite-embedding-278m-multilingual.png}
        
%         \label{fig:ibm}
%     \end{subfigure}

%     \caption{
%         Dimension–precision trade-off curves (nDCG@5) on NanoFeverRetrieval under 
%         1-bit, 1.58-bit, 2-bit, and INT4 quantization.
%     }
%     \vspace{-4ex}
%     \label{fig:summary}
% \end{figure}

% KaLM-Embedding/KaLM-embedding-multilingual-mini-instruct-v2.5

\begin{figure}[t!]
    \centering
    \includegraphics[width=.9\linewidth]{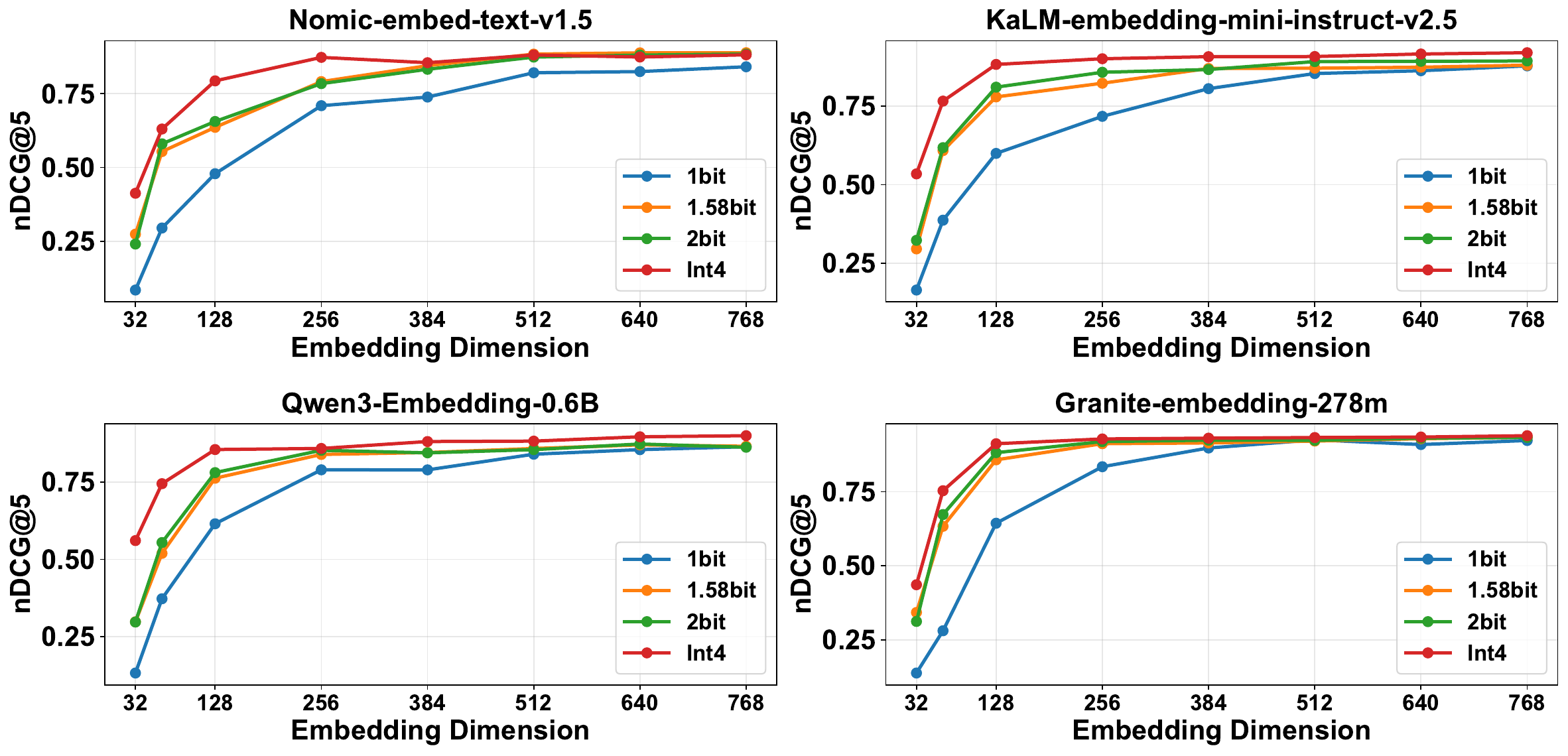}
    \vspace{-2ex}
    % \caption{Overview of the proposed hardware-aware representation learning framework. }
    % \caption{Retrieval accuracy curves for four embedding models across varying embedding dimensions under four quantization precisions (1-bit, 1.58-bit, 2-bit, and INT4), illustrating how performance changes as dimension and precision vary.}
        % The curves illustrate how performance scales as both dimension and precision are reduced.
    \caption{Retrieval accuracy curves for four embedding models across different embedding dimensions and four quantization precisions (1-bit, 1.58-bit, 2-bit, and INT4).}

    \vspace{-5ex}
    \label{fig:summary}
\end{figure}

To evaluate the generality of CQ-CiM, we include four embedding models on
% test four representative embedding models on 
NanoFeverRetrieval \cite{thorne-etal-2018-fever} benchmark: \textit{Nomic-embed-text-v1.5} \cite{nussbaum2024nomic}, \textit{KaLM-embedding-mini-instruct-v2.5} \cite{zhao2025kalmembeddingv2,hu2025kalmembedding}, \textit{Qwen3-Embedding-0.6B} \cite{qwen3embedding}, and \textit{Granite-embedding-278m} \cite{awasthy2025graniteembeddingmodels}. This selection covers models varying in scale and training objectives. For each model, embeddings are reduced to multiple target dimensions and quantized under four precision settings—1-bit, 1.58-bit, 2-bit, and Int4. We use nDCG@5 as the evaluation metric for retrieval.
% We report nDCG@5 as the primary metric to capture ranking accuracy and top-heavy retrieval behavior.

Across all models, Figure~\ref{fig:summary} shows a consistent trend: retrieval quality decreases smoothly as dimension is reduced, while 1.58-bit and 2-bit quantization preserve strong performance down to 128 dimensions, whereas 1-bit quantization degrades sharply at very low dimension. Although INT4 quantization achieves the strongest accuracy, low-bit quantization is far better suited to CiM. It avoids multi-step arithmetic and eliminates the shifting and stitching in higher-precision representation. In INT4, each value must be split across multiple CiM cells (e.g., four 1-bit SRAM cells), and the partial results must be shifted and stitched back together during accumulation. This process increases data movement, adds extra compute cycles, and amplifies noise from device mismatch. 

In contrast, CQ-CiM enables dot-product computation in a single crossbar step on modern multi-level NVM arrays operating at 2-bit quantization \cite{soliman2023first}. This reduces computational complexity and allows a substantially larger corpus under the same memory budget.

% We further report the training memory footprint and wall-clock time of each embedding model in Table~\ref{tab:training_vram_time}, illustrating the practical resource requirements associated with scaling CQ-CiM across heterogeneous encoders.
% We further assess the practical efficiency of CQ-CiM by measuring the end-to-end training time required to adapt each embedding model. As shown in Table~\ref{tab:training_vram_time}, our framework completes CiM-specific adaptation within only a few minutes across models of different scales, demonstrating that CQ-CiM's computational efficiency
% introduces minimal computational overhead and is easily scalable to heterogeneous encoders.

We further assess the practical efficiency of CQ-CiM by measuring the end-to-end training time required to adapt each embedding model. The results in Table~\ref{tab:training_vram_time} are obtained using a fixed target embedding dimension of 128D, matching the 2-bit CiM crossbar setting used throughout this work. As shown in the table, CQ-CiM completes CiM-specific adaptation within only a few minutes even for models with hundreds of millions of parameters, demonstrating that the framework incurs minimal computational overhead and scales efficiently across diverse embedding architectures.

\subsection{End-to-End RAG}

To close the loop between embedding quantization and system-level retrieval, we integrate device variation directly into an end-to-end RAG pipeline. Following the Gaussian conductance–variation model widely used in multi-level NVM characterization \cite{shim2020two}, and using the device parameters in Table~\ref{tab:var}, we derive a level-dependent $4\times4$ transition matrix that captures the probability of each 2-bit state flipping to another during readout. During evaluation, every quantized embedding is perturbed according to this transition matrix before retrieval.

We then evaluate the full RAG stack:  the embedding model (after noise injection) retrieves the top-5 documents for each query, and an LLM selects the single most relevant document among these candidates. The prediction is compared to the ground-truth document, producing an end-to-end accuracy metric that reflects both embedding robustness and LLM selection reliability.

\captionsetup[table]{position=bottom}
\begin{table}[t!]
\fontsize{8pt}{3pt}\selectfont
    \centering
    \small
    \vspace{-3ex}
    % \caption{Parameter size and end-to-end CQ-CiM training time for different embedding models (128D).}
    % \caption{Parameter size and end-to-end CQ-CiM training time for different embedding models at a fixed 128D target dimension and 2-bit CiM setting.}
    % \caption{\textcolor{red}{[TO DO]} Parameter size and end-to-end CQ-CiM training time at 128D and 2-bit CiM settings.}
    \caption{Parameter size and end-to-end CQ-CiM training time for different embedding models at a fixed 128D target dimension under a 2-bit CiM setting.}

    \resizebox{\columnwidth}{!}{

    \begin{tabular}{lcc}
        \toprule
        \textbf{Model} & \textbf{Para. Size} & \textbf{Time} \\
        \midrule
        Granite-embedding-278m      & 278\,M & 5m28s  \\
        Qwen3-Embedding-0.6B                     & 600\,M & 19m40s \\
        KaLM-embedding-mini-instruct-v2.5 & 500\,M & 14m18s \\
        Nomic-embed-text-v1.5                    & 137\,M & 4m53s  \\
        \bottomrule
    \end{tabular}
    % \caption{Training memory usage and time for different embedding models on 128 dimension.}
    }

    \vspace{-4ex}
    \label{tab:training_vram_time}
\end{table}

In these experiments, noise injection follows Eq.~\ref{eq:noise_injection} with $\epsilon=0.1$. LLMs in experiments operate in FP16 with temperature $=0$, and all embedding models are evaluated using batch size~16. For LoRA-enabled encoders, we set rank to 8, $\alpha$ to 16, and dropout to 0.05. 
% Experiments run on a NVIDIA A10.

As shown in Figure~\ref{fig:rag-llm-device}, the proposed compressed embeddings remain stable across datasets and LLM architectures even under realistic device-induced level-flip noise. Although device variation inevitably reduces retrieval performance, the degradation is modest, and the relative ordering among methods remains consistent. These results confirm that our low-bit representations are resilient not only in embedding-space metrics but also in a full RAG pipeline, where hardware noise affects both retrieval candidates and downstream LLM reasoning.

\section{Conclusion}
% In this paper, we introduce \textbf{CQ-CiM}, a unified, hardware-aware data shaping framework to solve the fundamental "representation gap" for RAG on CiM architectures. Our approach provides a solution to flexibly shape high-precision, high-dimension embeddings to fit the low-precision, low-dimension constraints of diverse CiM crossbar arrays. By jointly optimizing a LoRA adapter, a Compression head, and a Quantization head, our self-supervised framework learns to co-design the data representation for the hardware's physical properties. Experimental results show that CQ-CiM significantly outperforms traditional methods on diverse retrieval benchmarks and maintains high robustness against realistic hardware device noise. This paper marks the first work to propose a unified data shaping framework for comprehensive and robust CiM usage on RAG.

In this paper, we introduce \textbf{CQ-CiM}, a unified hardware-aware data-shaping framework that addresses the fundamental "representation gap" of deploying RAG on CiM architectures. Our approach flexibly transforms high-precision, high-dimension embeddings to meet the low-precision, low-dimension constraints of diverse CiM arrays. By jointly optimizing a LoRA adapter, a Compression head, and a Quantization head, 
our self-supervised framework learns embedding representations that are robust to device-level variation.
% our self-supervised framework co-designs data representations with CiM device variance. 
Experiments show that CQ-CiM outperforms traditional methods across retrieval benchmarks while maintaining strong robustness under realistic device noise. This is the first unified data shaping framework enabling comprehensive and reliable CiM-based RAG.

\clearpage
\bibliographystyle{unsrt}
\bibliography{reference, citations, citation_ICCAD_1, citation_ICCAD_2, bibliography_alp}
% \bibliography{citations}

\end{document}